\documentclass[conference,a4]{IEEEtran}

\usepackage{graphicx}
\graphicspath{{./figs/}}
\usepackage[T1]{fontenc}
\usepackage{amsmath,amssymb}
\usepackage{mathtools}
\usepackage{stackrel}
\usepackage{cite}
\usepackage{algorithmic}
\usepackage{url}
\usepackage{pgfplots}
\pgfplotsset{compat=1.18}
\usepackage{comment}
\usetikzlibrary{arrows.meta} 
\usepackage{threeparttable}
\usepackage{subcaption} 
\def\BibTeX{{\rm B\kern-.05em{\sc i\kern-.025em b}\kern-.08em
    T\kern-.1667em\lower.7ex\hbox{E}\kern-.125emX}}
\DeclareMathOperator*{\argmax}{arg\,max}

\begin{document}

\title{On the Connection Between Magnetic-Field Odometry Aided Inertial Navigation and Magnetic-Field SLAM}

\author{
\IEEEauthorblockN{1\textsuperscript{st} Isaac Skog}
\IEEEauthorblockA{School of Electrical Engineering\\ 
and Computer Science\\
KTH Royal Institute of Technology\\
Stockholm, Sweden\\
\url{skog@kth.se}}
\and
\IEEEauthorblockN{2\textsuperscript{nd} Manon Kok}
\IEEEauthorblockA{Delft Center for Systems and Control\\
Delft University of Technology\\
Delft, the Netherlands \\
\url{m.kok-1@tudelft.nl}}
\and
\IEEEauthorblockN{3\textsuperscript{rd} Gustaf Hendeby}
\IEEEauthorblockA{Dept. of Electrical Engineering\\
\textit{Link\"{o}ping University}\\
Link\"{o}ping, Sweden \\
\url{gustaf.hendeby@liu.se}}
\and
\IEEEauthorblockN{4\textsuperscript{th}  Chuan Huang}
\IEEEauthorblockA{School of Electrical Engineering 
and Computer Science\\
KTH Royal Institute of Technology\\
Stockholm, Sweden\\
\url{chuanh@kth.se}}
\and
\IEEEauthorblockN{5\textsuperscript{th} Thomas Edridge}
\IEEEauthorblockA{Delft Center for Systems and Control\\
Delft University of Technology\\
Delft, the Netherlands \\
\url{T.I.Edridge@tudelft.nl}}}

\maketitle

\begin{abstract}
Magnetic-field simultaneous localization and mapping (SLAM) using consumer-grade inertial and magnetometer sensors offers a scalable, cost-effective solution for indoor localization. However, the rapid error accumulation in the inertial navigation process limits the feasible exploratory phases of these systems. Advances in magnetometer array processing have demonstrated that odometry information, i.e., displacement and rotation information, can be extracted from local magnetic field variations and used to create magnetic-field odometry-aided inertial navigation systems. The error growth rate of these systems is significantly lower than that of standalone inertial navigation systems. This study seeks an answer to whether a magnetic-field SLAM system fed with measurements from a magnetometer array can indirectly extract odometry information --- without requiring algorithmic modifications --- and thus sustain longer exploratory phases. The theoretical analysis and simulation results show that such a system can extract odometry information and indirectly create a magnetic field odometry-aided inertial navigation system during the exploration phases.   However, practical challenges related to map resolution and computational complexity remain significant. 
\end{abstract}

\begin{IEEEkeywords}
SLAM, inertial navigation, magnetic field, odometry, sensor array  
\end{IEEEkeywords}

\IEEEpeerreviewmaketitle

\begin{figure}[t!]
    \centering
    \includegraphics[width=\linewidth]{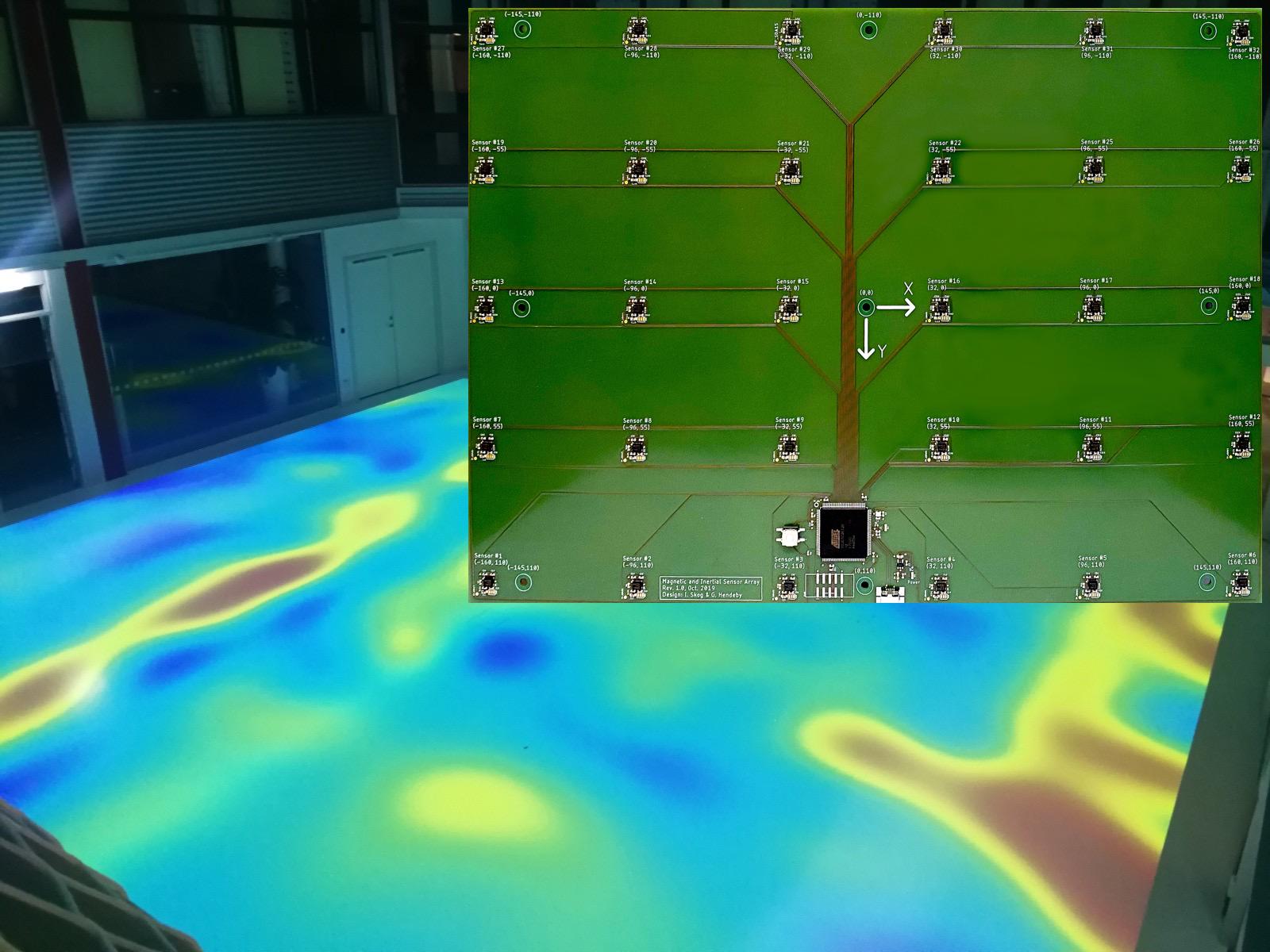}
    \caption{A magnetometer sensor array overloaded on a picture of the magnetic-field magnitude variations in a room. The array has 30 three-axis magnetometer sensors and one inertial measurement unit. The size of the array is $32 \times 22$ centimeters.}
    \label{fig:field_and_array}
\end{figure}

\section{Introduction}
Magnetic-field simultaneous localization and mapping (SLAM) systems constructed around consumer-grade inertial and magnetometer sensors offer a promising approach for developing scalable and cost-effective indoor localization systems~\cite{Kok2018MagSlam,manon2021MagneticField,robertsonFADJPKLB:2013,vallivaaraHKR:2011,coulinGGJF:2022}. The main challenge with the technology is the fast error accumulation in the inertial navigation process when using low-cost sensors, which means that only short exploration phases can be allowed. The faster the error in the inertial navigation process grows, the more frequently it is necessary to revisit previously mapped areas and exploit the earlier learned magnetic-field map to reduce the navigation error. Short exploration phases mean that only small areas can be mapped before revisiting previously mapped regions, severely limiting the system's practical use.

Although the error accumulation in inertial navigation can be limited using additional information from, e.g., zero velocity updates from a sensor placed on the foot of the user~\cite{Wahlstrom2021Fifteen} or motion constraints~\cite{Engelsman2023}, the rapid error accumulation during the exploration phases remains one of the key limitations on the practical usage of magnetic-field SLAM. Recently, a separate line of research on magnetometer sensor array processing has demonstrated that odometry information, i.e., displacement and orientation change information, can be extracted from local magnetic field variations and used to aid an inertial navigation system~\cite{Chesneau2017ImprovingMagnetoInertial,zmitri2020magnetic,huang2023mains}. The result is a navigation system with a significantly slower error growth rate than a standalone inertial navigation system without any constraints regarding its motion or position on the user's body.  

This paper investigates the connection between magnetic-field odometry-aided inertial navigation and magnetic-field SLAM. More precisely, it seeks an answer to whether existing magnetic-field SLAM methods can indirectly extract odometry information when fed data from a magnetometer sensor array, similar to that in Fig.~\ref{fig:field_and_array}. To that end, the paper presents: (i) a theoretical analysis of the connections between magnetic-field odometry-aided inertial navigation and magnetic-field SLAM; (ii) a simulation study that verifies the theoretical findings; and (iii) a discussion about the practical challenges related to the magnet-field map resolution in the two types of systems.


\section{General System Model}
Consider a navigation platform equipped with a sensor assembly similar to that in Fig.~\ref{fig:field_and_array} consisting of $n_y$ three-axis magnetometers and one inertial measurement unit. Let $y^{(i)}_\tau$ denote the measurement from the $i$:th magnetometer at time instant $\tau$, and let $y_\tau\triangleq \{y^{(i)}_\tau\}_{i=1}^{n_y}$ denote the set of all magnetometer measurements at time instant $\tau$. Further, let $u_\tau$ denote the measurements from the inertial measurement unit\footnote{The forthcoming analysis also holds for measurements from other types of dead-reckoning sensors, such as wheel encoders or Doppler velocity loggers.}. Moreover, let 
\begin{equation}
    x_\tau=\begin{bmatrix}
        r_\tau\\
        q_\tau\\
        \xi_\tau
    \end{bmatrix}
\end{equation}
denote the navigation state of the platform, which is to be inferred. Here, $r_\tau$ and $q_\tau$ denote the platform's location and orientation. Further, $\xi_\tau$ denotes the auxiliary state needed to model the platform dynamics and sensor behavior.  

Next, let $m(r)\in\mathbb{R}^{3}$ denote the ambient magnetic field at location $r$, which is modeled as a continuous stochastic process with the probability density function (PDF) 
\begin{equation}
     p(m(r_1), m(r_2),\ldots,m(r_N)),   
\end{equation}
where $r_1,r_2,\ldots,r_N$ are arbitrarily locations. Here, it has been explicitly highlighted that the magnetic field depends on the location $r$, but in the sequel, this will not be done for notation brevity. Commonly, the stochastic process is assumed to be a Gaussian process as it supports analytical inference and encoding of physical properties about the magnetic field, such as curl- and divergence-freeness~\cite{Kok2018MagSlam,Niklas2013Modeling}.      

Given the defined quantities, let the following Markov model describe the dynamics and observation of the navigation platform    
\begin{subequations}\label{eq:general model}
\begin{align}
        &p(x_{\tau+1}|x_\tau,u_\tau)\\
        &p(y_\tau|x_\tau,m)
\end{align}
\end{subequations}
where it is assumed that measurements from the magnetometer are independent so that
\begin{equation}\label{eq:multiple magnetometers}
   p(y_\tau|x_\tau,m)=\prod_{i=1}^{n_y}p(y^{(i)}_\tau|x_\tau,m).
\end{equation}
Here $p(a|b)$ denotes the conditional PDF of $a$ given $b$. This general system model encases most models used for aided inertial navigation and SLAM systems. This includes both the magnetic-field odometry-aided inertial navigation systems presented in~\cite{Chesneau2017ImprovingMagnetoInertial,Vissiere2007magnetometers,dorveaux2011combining,dorveaux2011presentation,Chuan2022Magnetic,chesneau2016motion,zmitri2020magnetic,zmitri2022BILSTM,huang2023mains}, where $n_y>1$, as well as the magnetic-field SLAM systems presented in~\cite{manon2021MagneticField,Viset2022EKF,Kok2018MagSlam,lee2020magslam}, where $n_y=1$.    
\section{Magnetic-Field Odometry Aided Inertial Navigation versus Magnetic-Field SLAM}\label{s:odometry vs SLAM}
Next, the connection between magnetic-field odometry-aided inertial navigation and magnetic-field SLAM will be highlighted by examining the Bayesian filter recursions for these two types of systems. 

\subsection{Filter Recursions For Magnetic-Field SLAM}
Let $y_{1:t}\triangleq\{y_\tau\}_{\tau=1}^t$, $x_{1:t}\triangleq\{x_\tau\}_{\tau=1}^t$, and $u_{1:t}\triangleq\{u_\tau\}_{\tau=1}^t$. The goal of the magnetic-field SLAM system is to calculate the posterior PDF
\begin{equation}\label{eq:posterior pdf slam}
p(x_{t},m|y_{1:t},u_{1:t-1}) \quad\text{or}\quad p(x_{1:t},m|y_{1:t},u_{1:t-1}). 
\end{equation}
Extended Kalman filter-based SLAM algorithms generally calculate the former posterior PDF, whereas smoothing and Rao-Blackwellised particle filter-based algorithms calculate the latter posterior PDF~\cite{DurrantWhyte2006,Kok_Solin_Schön_2024}. Without loss of generality, the forthcoming discussion will only be about the calculation of $p(x_{t},m|y_{1:t},u_{1:t-1})$, which can be recursively calculated via the time and measurement updates
\begin{subequations}
\begin{multline}
      p(x_{t+1}, m | y_{1:t},u_{1:t}) = \int p(x_{t+1} | x_{t}, u_{t})\\ \times p(x_{t}, m | y_{1:t},u_{1:t-1}) \, dx_{t},
\end{multline}
and
\begin{multline}
p(x_{t}, m | y_{1:t},u_{1:t-1}) = \frac{p(y_t | x_t, m) p(x_{t}, m | y_{1:t-1},u_{1:t-1})}{p(y_t | y_{1:t-1},u_{1:t-1})}\\
=\frac{p(x_{t}, m | y_{1:t-1},u_{1:t-1})\prod_{i=1}^{n_y}p(y^{(i)}_t|x_t,m)}{p(y_t | y_{1:t-1},u_{1:t-1})},
\label{eq:posteriorCalculation}
\end{multline}
\end{subequations}
respectively. Here, $p(y_t | y_{1:t-1},u_{1:t-1})$ is the data evidence, which can be calculated via marginalization.     

From the posterior PDF, the minimum mean square estimate of the navigation state $x_t$ can be calculated as
\begin{equation}
    \hat{x}_t=\iint x_t p(x_{t},m|y_{1:t}) d x_t d m
\end{equation}
and its covariance as
\begin{equation}
    \text{Cov}\{\hat{x}_t\}=\iint (x_t-\hat{x}_t)(x_t-\hat{x}_t)^\top p(x_{t},m|y_{1:t}) d x_t dm.
\end{equation}

Noteworthy is that the only difference between a traditional magnetic-field SLAM implementation with $n_y=1$ and an implementation that uses an array of magnetometers, i.e., $n_y>1$, is that in the latter case, the measurement update includes measurements from multiple spatially distributed magnetometers.       

\subsection{Filter Recursions For Magnetic-Field Odometry Aided Inertial Navigation}
The goal of the magnetic-field odometry-aided INS is to calculate the posterior PDF
\begin{equation}\label{eq:odo pdf}
p(x_{t}|y_{1:t},u_{1:t-1}).
\end{equation}
That is, the aim is to calculate only the conditional distribution of the state $x_t$, and not the magnetic field $m$. Typical reasons for not wanting to calculate the magnetic field $m$ are (a) computational and memory constraints relating to inferring and storing a model of the magnetic field or (b) that the magnetic field is time-varying and can only be considered stationary over a limited period. 

The posterior PDF in \eqref{eq:odo pdf} can be calculated from the SLAM posterior PDF via marginalization, i.e.,
\begin{equation}\label{eq:marginalization m}
    p(x_{t}|y_{1:t},u_{1:t-1})=\int p(x_{t},m|y_{1:t},u_{1:t-1})\,dm.
\end{equation}    
Hence, from a theoretical perspective, the existing magnetic-field SLAM systems can extract magnetic-field odometry information if measurements are supplied from a magnetometer array. More importantly, as will be shown next, without further assumption about the behavior of the magnetic-field $m(r)$ and the trajectory $x_{1:t}$, the posterior PDF $p(x_{t}|y_{1:t},u_{1:t-1})$ can only be calculated from the SLAM posterior PDF. This can be seen by rewriting $p(x_{t}|y_{1:t},u_{1:t-1})$ using \eqref{eq:marginalization m} and then applying Bayes' theorem. This gives
\begin{multline}\label{eq:odo meas update}
    p(x_{t}|y_{1:t},u_{1:t-1})=\int p(x_{t},m|y_{1:t},u_{1:t-1}) dm\\
    \propto \int p(y_t|x_{t},m,y_{1:t-1},u_{1:t-1})\\ \times p(x_{t},m|y_{1:t-1},u_{1:t-1})dm\\
    =\int\!p(y_t|x_{t},m)p(x_{t},m|y_{1:t-1},u_{1:t-1}) dm
\end{multline}
where
\begin{multline}\label{eq:odo time update}
    p(x_{t},m|y_{1:t-1},u_{1:t-1})=\int p(x_t|x_{t-1},u_{t-1}) \\
    \times p(x_{t-1},m|y_{1:t-1},u_{1:t-2})d x_{t-1}. 
\end{multline}
From \eqref{eq:odo meas update} and \eqref{eq:odo time update} it is clear that to calculate the posterior PDF $p(x_{t}|y_{1:t},u_{1:t-1})$ the SLAM posterior PDF $p(x_{t-1},m|y_{1:t-1},u_{1:t-2})$ is needed. This implies that without further assumptions about the behavior of the magnetic field $m(r)$ and the trajectory $x_{1:t}$ existing magnetic-field odometry-aided inertial navigation system solutions, such as \cite{huang2023mains}, will only calculate an approximation of $p(x_{t}|y_{1:t},u_{1:t-1})$. Hence, they will provide suboptimal navigation performance. 

The need to calculate the full SLAM posterior PDF to find the posterior PDF for the magnetic-field odometry-aided inertial navigation can be relaxed by introducing the following assumptions: (i) the system only operates in an exploration phase; (ii) information about the magnetic field at faraway locations provides no information about the field at the current location; and (iii) the inertial measurements $u_{1:t}$ are such that $x_t$ is uni-modal distributed. 


\section{Implementation Challenges}

\begin{figure*}[h!]
    \centering
    \includegraphics[width=\textwidth]{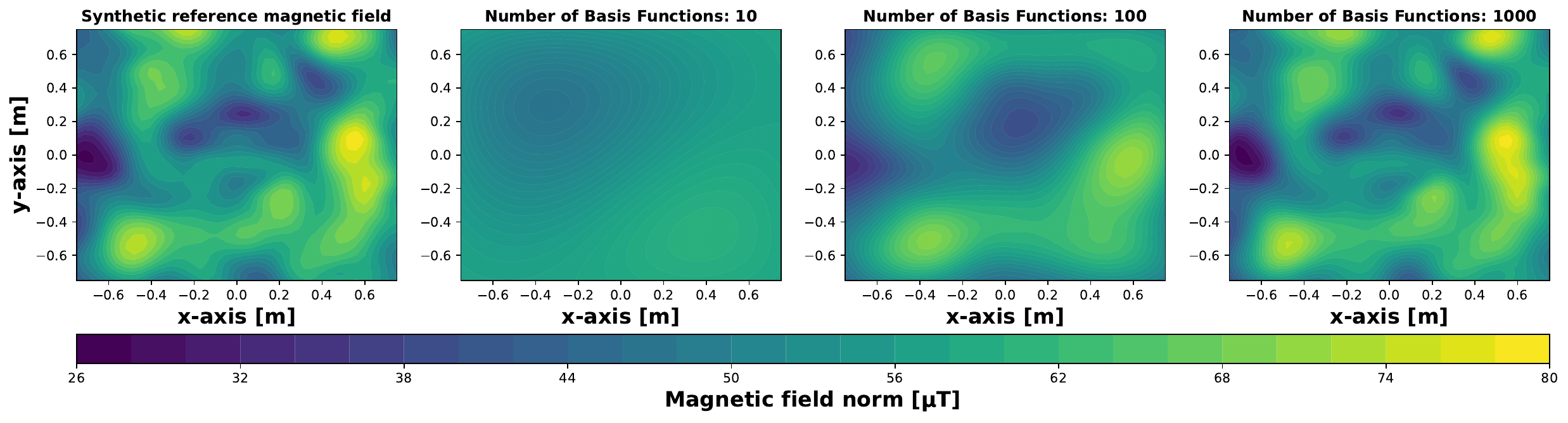}
    \caption{Illustration of the magnetic field model resolution for different numbers of basis functions. A synthetic reference field (left) and magnetic field models with 10, 100, and 1000 basis functions fitted to the reference field are shown. Normalized by the area, this corresponds to approximately 5, 50, and 500 basis functions per square meter.}
    \label{fig:resolutionNumberOfBasisFunctions}
\end{figure*}

Although a magnetic-field SLAM system can theoretically extract odometry information if it is fed measurements from a magnetometer array, several aspects make it practically challenging to implement such a system. A key challenge is how the SLAM system should model the ambient magnetic field  $m(r)$ so that odometry information can be extracted while maintaining a reasonable computational complexity. Current magnetic-field SLAM systems exploit the larger-scale variations in the ambient magnetic field and model field variations on a length scale of a meter~\cite{Viset2022EKF}. In contrast, a magnetic-field odometry-aided inertial navigation system exploits the magnetic field variations at the scale of the array to extract odometry information~\cite{huang2023mains, skog2021magnetic}. Hence, a magnetic-field model that can represent the field variations on a sub-meter length scale is needed for a system that uses an array similar to that in Fig.~\ref{fig:field_and_array}. Further, the resolution of the magnetometers used in the system must be such that they can reliably resolve these fine-grained variations in the magnetic field across the array.   

Two commonly used ways of modeling the magnetic field in SLAM and magnetic-field odometry systems are (approximate) Gaussian process models~\cite{Kok2018MagSlam} and polynomial models~\cite{skog2021magnetic}. Both models approximate the magnetic field in terms of a basis function expansion, i.e.,   
\begin{equation}\label{eq:basis function expansion}
    m(r)\approx \sum_{i=1}^{n_b} \phi_i(r)\theta_i
\end{equation}
where $\phi_i(r)$ and  $\theta_i$ denote the $i$:th basis function and basis function weight, respectively. Further, $n_b$ is the total number of basis functions. The computational complexity of estimating these weights given $t$ magnetic field measurements is $O(t\,n^2_b)$~\cite{solinS:2020}. In Fig.~\ref{fig:resolutionNumberOfBasisFunctions}, the model resolution for different numbers of basis functions is illustrated. As can be seen, for the magnetic-field model to accurately capture variations on a sub-meter length scale, between 50 to 500 basis functions per square meter are needed. In contrast, 5 to 50 basis functions per square meter are needed if variations on a meter scale are to be captured. Hence, if a magnetic-field SLAM system is to extract odometry information, the resolution of the magnetic-field model, a.k.a magnetic-field map, must be increased by an order of magnitude. This corresponds to a computational complexity increase of two orders of magnitude if no algorithmic changes are made beyond the change in the number of basis functions in the magnetic field model.

\begin{figure}[h!]
    \centering
        \begin{subfigure}{\columnwidth}
        \centering
        \includegraphics[trim={0cm 0cm 0cm 0.6cm}, clip,width=\columnwidth]{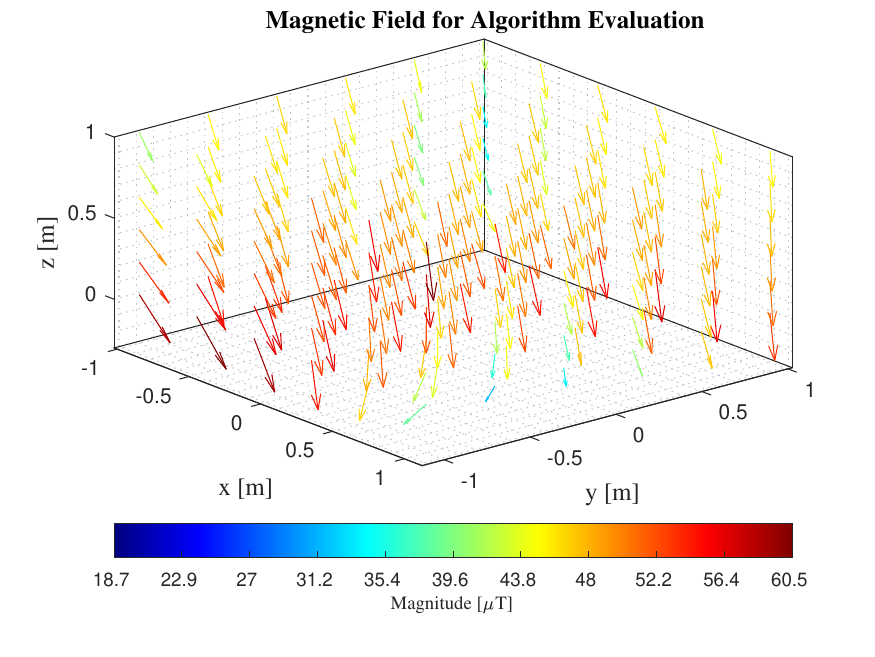}
        \caption{Illustration of the magnetic field in the simulation. The magnetic field model is based on real measurements, to which a sum of magnetic-dipole models has been fitted.}
        \label{fig:magnetic-field}
    \end{subfigure}
    \begin{subfigure}{\columnwidth}
        \centering
        \includegraphics[trim={1.5cm 6.5cm 2cm 5.5cm}, clip, width=\columnwidth]{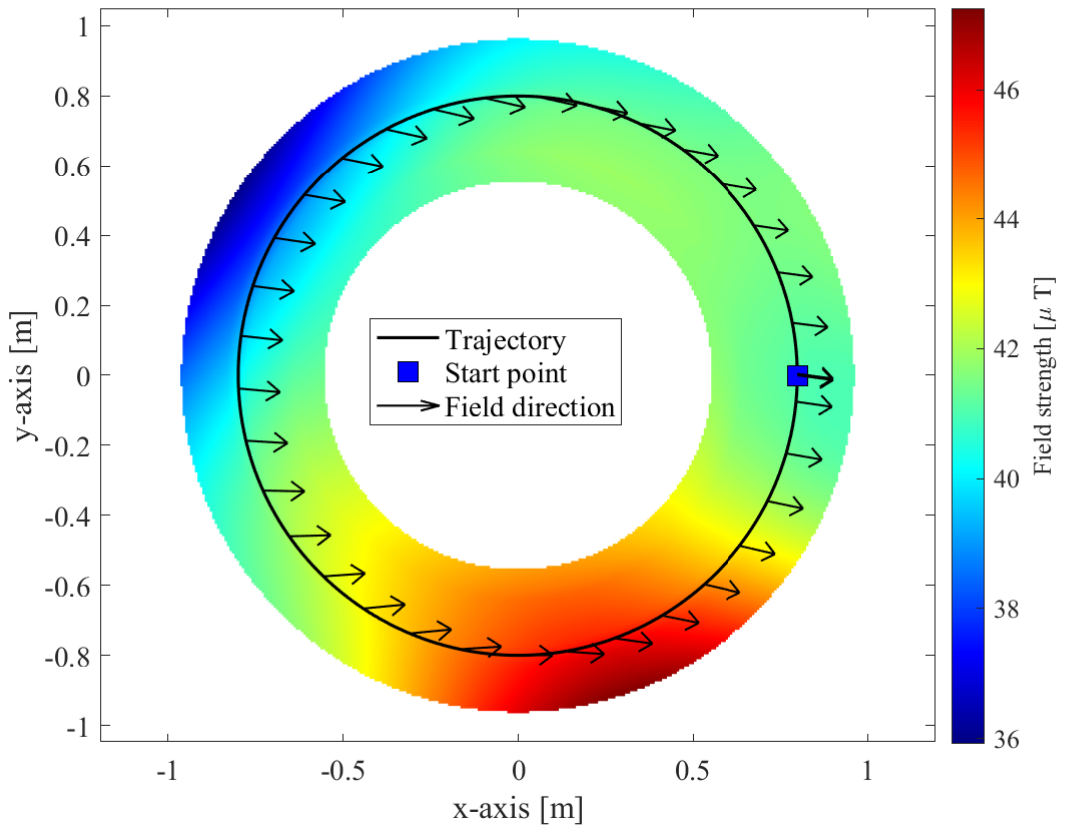}
        \caption{Illustration of the circular trajectory used in the simulation. The platform moves counterclockwise. Also shown are the magnetic-field magnitude variations in the neighborhood of the trajectory and the direction of the in-plane component of the magnetic field.}
        \label{fig:trajectory}
    \end{subfigure}
    \caption{Illustration of (a) the magnetic field and (b) the circular trajectory used in the simulations.}
    \label{fig:simulation-setup}
\end{figure}

\section{Simulation Example}
The following simulation was performed to verify that a magnetic-field SLAM system can extract odometry information if provided with measurements from an array, not only in theory but also in practice.

\subsection{Simulation Setup}
The magnetic field shown in Fig.~\ref{fig:magnetic-field} was simulated. The simulated magnetic field was generated by fitting a sum of magnetic dipole models to real-world data collected in a volume of approximately $4$\,m$^3$ in the room shown in Fig. \ref{fig:field_and_array}. Then, measurements from a sensor array moving in a circular trajectory, as shown in Fig.~\ref{fig:trajectory}, were generated. The sensor array was similar to that in Fig.~\ref{fig:field_and_array} and moved counterclockwise at an angular velocity of $30\,^\circ/\mathrm{s}$ for multiple laps. The inertial measurements were simulated by adding biases and white Gaussian noise to the true specific force and angular velocity readings. The biases were modeled as random walk processes as suggested in~\cite{Farrell2022Inertial}. The magnetic field measurements from the magnetometers in the array were simulated by adding white Gaussian noise to the true magnetic field readings. All sensor parameters are summarized in Tab.~\ref{T:settings}. A second dataset was also simulated to show that a single low noise magnetometer cannot replace the spatially distributed magnetometers in the array. In the second dataset, the inertial measurements were kept the same, but measurements from a single magnetometer with a noise density of $\frac{1}{30}$ of the value in Tab.~\ref{T:settings} were generated.

A magnetic-field SLAM system was implemented using the Georgia Tech Smoothing and Mapping (GTSAM) toolbox~\cite{gtsam}. A basis function expansion model as that in \eqref{eq:basis function expansion} and with basis functions similar to those in \cite{Viset2022EKF} was used as the magnetic-field model. A total of $n_b=250$ basis functions (corresponding to 75 basis functions per square meter) were used. The implemented magnetic-field SLAM system calculates the maximum a posteriori estimation of the navigation state, i.e., it calculates the estimate
\begin{equation}
    \hat{x}_{t}=\argmax_{x_{t}}\bigl(\,\,\max_{x_{1:t-1},m} p(x_{1:t},m|y_{1:t},u_{1:t-1})\bigr)
\end{equation}
The implemented magnetic-field SLAM system was used to process the two generated data sets. The dataset with measurements from the simulated array was also processed using the magnetic-field odometry-aided inertial navigation system algorithm presented in~\cite{huang2023mains}. 

\subsection{Results}
\begin{figure*}[h!]
    \centering
    \includegraphics[trim={2.8cm 0.5cm 3cm 0cm}, clip,width=\textwidth]{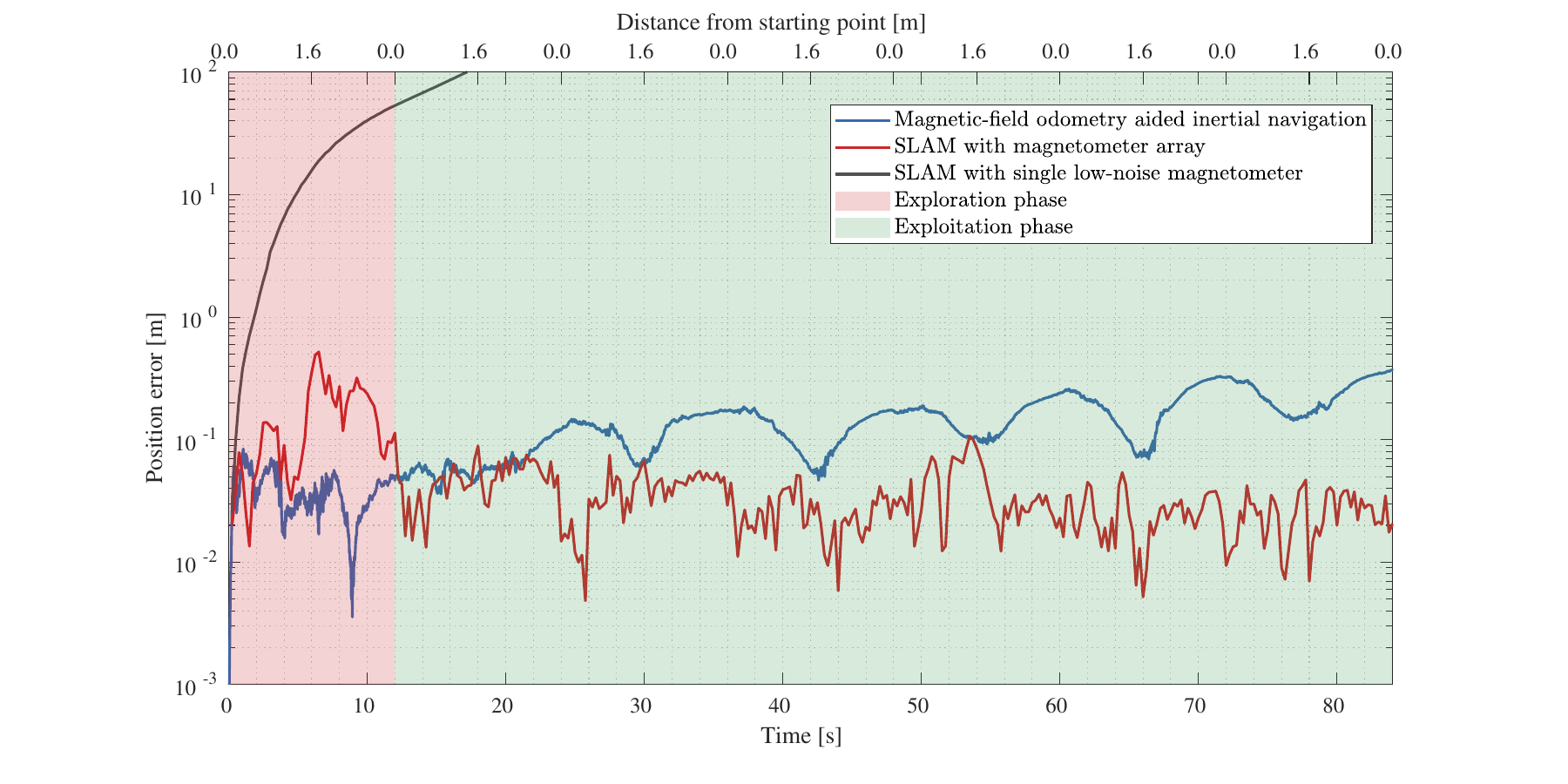}
    \caption{Position errors of the magnetic-field SLAM system when processing the two datasets. The first dataset includes measurements from an array (red line), and the second dataset only includes measurements from a single low-noise magnetometer (black line). Also shown is the position error of the magnetic-field odometry-aided inertial navigation system in~\cite{huang2023mains} (blue line).}
    \label{fig:position error}
\end{figure*}

In Fig.~\ref{fig:position error}, the position error of the implemented magnetic-field SLAM system when processing the two datasets is shown. Also shown is the position error of the magnetic-field odometry-aided inertial navigation system. From the following things can be observed. When processing the data from the simulated magnetometer array, the magnetic-field SLAM system can extract odometry information and reduce error growth during the exploration phase by two orders of magnitude compared to when magnetometer data from a single low-cost magnetometer is processed. Further, as expected, the SLAM system has a limited position error, whereas the magnetic-field odometry-aided inertial navigation system has an error that grows with time; no loop-closure information is gained when revisiting previous locations.

\begin{table}[tb!]
  \centering
  \caption{Sensor parameters used in the simulation}\label{T:settings}
  \begin{tabular}{lll}
  \hline
  \hline
  \textbf{Description} & \textbf{Value} & \textbf{Unit} \\
  \hline
  Accelerometer noise density & 0.01 & m/s$^2$/$\sqrt{\text{Hz}}$ \\
  Accelerometer bias & [$-0.32$, $-0.59$, $-0.37$]$^\top$&  m/s$^2$\\
  Acceleration random walk rate& 1.0$\times$10$^{-6}$ & m/s$^{\frac{5}{2}}$ \\
  Gyroscope noise density & 0.05  & ${}^\circ/s/\sqrt{\text{Hz}}$ \\
  Gyroscope bias & [$-0.01$, $-1.39$, $-2.14$]$^\top$&  ${}^\circ/s$\\
  Angle random walk rate& 1.0$\times$10$^{-5}$ & ${}^\circ/s^{\frac{3}{2}}$ \\
  Magnetometer noise density & 0.02 & $\mu \text{T}/\sqrt{\text{Hz}}$ \\
  Sensor sampling rate & 100 & Hz\\
  \hline
  \hline
  \end{tabular}
\end{table}

\section{Conclusion}
Both the theoretical and simulation results show that a magnetic-field SLAM system fed with measurements from a magnetometer array can indirectly extract odometry information without requiring algorithmic modifications and thus sustain longer exploratory phases. The theoretical results also show that, without further assumptions about the magnetic field and the system's trajectory, an optimal magnetic-field odometry-aided inertial navigation system can only be realized by solving the SLAM problem. The former is significant as it lays the foundation for developing low-cost magnetic field SLAM systems for large-scale indoor environments. However, the computational complexity challenges originating from the need to use a high-resolution magnetic field map to extract odometry information remain significant. Future research should investigate ways to create and use multi-resolution maps within magnetic-field SLAM systems. Especially the concept of using a local high-resolution map, as in magnetic-field odometry-aided inertial navigation systems, in conjugate with a global low-resolution map, as in current magnetic-field SLAM systems, should be explored.

\section*{Acknowledgment}
This work has been partially funded by the Swedish Research Council project 2020-04253 \emph{Tensor-field based localization}, the Vinnova, Sweden’s Innovation Agency, and Swedish Armed Forces, project  2024-03194~\emph{Electromagnetic navigation for smaller unmanned underwater vehicles}, and the Dutch Research Council (NWO) research program Veni project 18213 \emph{Sensor Fusion For Indoor Localisation Using The Magnetic Field} and by the Sensor AI Lab, under the AI Labs program of Delft University of Technology.

\IEEEtriggeratref{19}
\bibliographystyle{IEEEtran}
\bibliography{IEEEabrv,ref}

\begin{thebibliography}{10}
\providecommand{\url}[1]{#1}
\csname url@samestyle\endcsname
\providecommand{\newblock}{\relax}
\providecommand{\bibinfo}[2]{#2}
\providecommand{\BIBentrySTDinterwordspacing}{\spaceskip=0pt\relax}
\providecommand{\BIBentryALTinterwordstretchfactor}{4}
\providecommand{\BIBentryALTinterwordspacing}{\spaceskip=\fontdimen2\font plus
\BIBentryALTinterwordstretchfactor\fontdimen3\font minus
  \fontdimen4\font\relax}
\providecommand{\BIBforeignlanguage}[2]{{%
\expandafter\ifx\csname l@#1\endcsname\relax
\typeout{** WARNING: IEEEtran.bst: No hyphenation pattern has been}%
\typeout{** loaded for the language `#1'. Using the pattern for}%
\typeout{** the default language instead.}%
\else
\language=\csname l@#1\endcsname
\fi
#2}}
\providecommand{\BIBdecl}{\relax}
\BIBdecl

\bibitem{Kok2018MagSlam}
M.~Kok and A.~Solin, ``Scalable magnetic field {SLAM} in 3{D} using {G}aussian
  process maps,'' in \emph{Proc. Int. Conf. on Information Fusion}, Cambridge,
  United Kingdom, July 2018, pp. 1353--1360.

\bibitem{manon2021MagneticField}
F.~Viset, J.~T. Gravdahl, and M.~Kok, ``{Magnetic field norm {SLAM} using
  {G}aussian process regression in foot-mounted sensors},'' in \emph{European
  Control Conference}, Rotterdam, Netherlands, June 2021, pp. 392--398.

\bibitem{robertsonFADJPKLB:2013}
P.~Robertson, M.~Frassl, M.~Angermann, M.~Doniec, B.~J. Julian,
  M.~Garcia~Puyol, M.~Khider, M.~Lichtenstern, and L.~Bruno, ``Simultaneous
  localization and mapping for pedestrians using distortions of the local
  magnetic field intensity in large indoor environments,'' in \emph{Proceedings
  of the IEEE International Conference on Indoor Positioning and Indoor
  Navigation (IPIN)}, Montb{\'e}liard, France, Oct. 2013, pp. 1--10.

\bibitem{vallivaaraHKR:2011}
I.~Vallivaara, J.~Haverinen, A.~Kemppainen, and J.~Roning, ``Magnetic
  field-based {SLAM} method for solving the localization problem in mobile
  robot floor-cleaning task,'' in \emph{Proc. Int. Conf. on Advanced Robotics},
  Tallinn, Estonia, June 2011, pp. 198--203.

\bibitem{coulinGGJF:2022}
J.~Coulin, R.~Guillemard, V.~Gay-Bellile, C.~Joly, and A.~de~La~Fortelle,
  ``Online magnetometer calibration in indoor environments for magnetic
  field-based {SLAM},'' in \emph{Proc. Int. Conf. on Indoor Positioning and
  Indoor Navigation}, 2022.

\bibitem{Wahlstrom2021Fifteen}
J.~Wahlström and I.~Skog, ``Fifteen years of progress at zero velocity: A
  review,'' \emph{IEEE Sensors Journal}, vol.~21, no.~2, pp. 1139--1151, 2021.

\bibitem{Engelsman2023}
D.~Engelsman and I.~Klein, ``Information-aided inertial navigation: A review,''
  \emph{IEEE Trans. on Inst. and Meas.}, vol.~72, pp. 1--18, Nov. 2023.

\bibitem{Chesneau2017ImprovingMagnetoInertial}
C.-I. Chesneau, M.~Hillion, J.-F. Hullo, G.~Thibault, and C.~Prieur,
  ``Improving magneto-inertial attitude and position estimation by means of a
  magnetic heading observer,'' in \emph{Proc. Int. Conf. on Indoor Positioning
  and Indoor Navigation}, Sapporo, Japan, Sep. 2017.

\bibitem{zmitri2020magnetic}
M.~Zmitri, H.~Fourati, and C.~Prieur, ``{Magnetic Field Gradient-Based EKF for
  Velocity Estimation in Indoor Navigation},'' \emph{Sensors}, vol.~20, no.~20,
  p. 5726, 2020.

\bibitem{huang2023mains}
C.~Huang, G.~Hendeby, H.~Fourati, C.~Prieur, and I.~Skog, ``{MAINS}: A
  magnetic-field-aided inertial navigation system for indoor positioning,''
  \emph{{IEEE} Sensors J.}, vol.~24, no.~9, pp. 15\,156--15\,166, 2024.

\bibitem{Niklas2013Modeling}
N.~Wahlström, M.~Kok, T.~B. Schön, and F.~Gustafsson, ``Modeling magnetic
  fields using gaussian processes,'' in \emph{2013 IEEE International
  Conference on Acoustics, Speech and Signal Processing}, 2013, pp. 3522--3526.

\bibitem{Vissiere2007magnetometers}
D.~Vissiere, A.~Martin, and N.~Petit, ``Using distributed magnetometers to
  increase imu-based velocity estimation into perturbed area,'' in \emph{Proc.
  IEEE Conference on Decision and Control}, New Orleans, LA, USA, Dec. 2007,
  pp. 4924--4931.

\bibitem{dorveaux2011combining}
E.~Dorveaux, T.~Boudot, M.~Hillion, and N.~Petit, ``Combining inertial
  measurements and distributed magnetometry for motion estimation,'' in
  \emph{Proc. American Control Conference}, San Francisco, CA, USA, June 2011,
  pp. 4249--4256.

\bibitem{dorveaux2011presentation}
E.~Dorveaux and N.~Petit, ``Presentation of a magneto-inertial positioning
  system: navigating through magnetic disturbances,'' in \emph{Int. Conf. on
  Indoor Positioning and Indoor Navigation}, Guimaraes, Portugal, Sep. 2011.

\bibitem{Chuan2022Magnetic}
C.~Huang, G.~Hendeby, and I.~Skog, ``A tightly-integrated magnetic-field aided
  inertial navigation system,'' in \emph{Proc. Int. Conf. on Information
  Fusion}, Linköping, Sweden, July 2022, pp. 1--8.

\bibitem{chesneau2016motion}
C.-I. Chesneau, M.~Hillion, and C.~Prieur, ``{Motion estimation of a rigid body
  with an EKF using magneto-inertial measurements},'' in \emph{Int. Conf. on
  Indoor Positioning and Indoor Navigation}, Alcalá de Henares, Spain, Oct.
  2016.

\bibitem{zmitri2022BILSTM}
M.~Zmitri, H.~Fourati, and C.~Prieur, ``Bi{LSTM} network-based extended kalman
  filter for magnetic field gradient aided indoor navigation,'' \emph{{IEEE}
  Sensors J.}, vol.~22, no.~6, pp. 4781--4789, 2022.

\bibitem{Viset2022EKF}
F.~Viset, R.~Helmons, and M.~Kok, ``An extended {K}alman filter for magnetic
  field slam using gaussian process regression,'' \emph{Sensors}, vol.~22,
  no.~8, 2022.

\bibitem{lee2020magslam}
T.~N. Lee and A.~J. Canciani, ``Magslam: Aerial simultaneous localization and
  mapping using earth's magnetic anomaly field,'' \emph{Navigation}, vol.~67,
  no.~1, pp. 95--107, 2020.

\bibitem{DurrantWhyte2006}
H.~Durrant-Whyte and T.~Bailey, ``Simultaneous localization and mapping: part
  i,'' \emph{{IEEE} Robot. Autom. Mag.}, vol.~13, no.~2, pp. 99--110, Jun.
  2006.

\bibitem{Kok_Solin_Schön_2024}
M.~Kok, A.~Solin, and T.~B. Schön, ``Rao-{B}lackwellized particle smoothing
  for simultaneous localization and mapping,'' \emph{Data-Centric Engineering},
  vol.~5, p. e15, 2024.

\bibitem{skog2021magnetic}
I.~Skog, G.~Hendeby, and F.~Trulsson, ``Magnetic-field based odometry – an
  optical flow inspired approach,'' in \emph{Int. Conf. on Indoor Positioning
  and Indoor Navigation}, Lloret de Mar, Spain, Nov. 2021.

\bibitem{solinS:2020}
A.~Solin and S.~S{\"a}rkk{\"a}, ``Hilbert space methods for reduced-rank
  {G}aussian process regression,'' \emph{Statistics and Computing}, vol.~30,
  pp. 419--446, Mar. 2020.

\bibitem{Farrell2022Inertial}
J.~A. Farrell, F.~O. Silva, F.~Rahman, and J.~Wendel, ``Inertial measurement
  unit error modeling tutorial: Inertial navigation system state estimation
  with real-time sensor calibration,'' \emph{IEEE Control Systems Magazine},
  vol.~42, no.~6, pp. 40--66, 2022.

\bibitem{gtsam}
\BIBentryALTinterwordspacing
F.~Dellaert and G.~Contributors, ``borglab/gtsam,'' May 2022. [Online].
  Available: \url{https://github.com/borglab/gtsam)}
\BIBentrySTDinterwordspacing

\end{thebibliography}

\end{document}